\documentclass[prl]{revtex4-2}

\usepackage{xr}

\usepackage{graphicx}
\usepackage{times}
\usepackage[hyphenbreaks]{breakurl}
\usepackage{amsmath}
\usepackage{amssymb}

\usepackage{hyperref}
\usepackage{cleveref}

\newcommand{\MSun}{\mbox{${M}_\odot$}}
\newcommand{\Msun}{\mbox{${M}_\odot$}}
\newcommand{\RSun}{\mbox{${R}_\odot$}}

\newcommand{\tAa}{\mbox{t$_{11}$}}
\newcommand{\tAb}{\mbox{t$_{12}$}}
\newcommand{\tB}{\mbox{t$_{2}$}}
\newcommand{\pA}{\mbox{p$_{1}$}}
\newcommand{\pB}{\mbox{p$_{2}$}}

\renewcommand{\thesection}{\arabic{section}}
\renewcommand{\thesubsection}{\thesection.\arabic{subsection}}
\def\apgt{\ {\raise-.5ex\hbox{$\buildrel>\over\sim$}}\ }
\def\aplt{\ {\raise-.5ex\hbox{$\buildrel<\over\sim$}}\ }
\def\lteq{\ {\raise-.5ex\hbox{$\buildrel<\over-$}}\ }
\def\gteq{\ {\raise-.5ex\hbox{$\buildrel>\over-$}}\ }

\begin{document}

\title{The origin of the most recently ejected OB runaway star from the R136 cluster}

\author{Simon Portegies Zwart}
\affiliation{Leiden Observatory, Leiden University, PO Box 9513, 2300 RA, Leiden, The Netherlands.}
\email{spz@astronomy.nl}
\author{Mitchel Stoop}
\affiliation{Anton Pannekoek Institute for Astronomy, University of Amsterdam, Science Park 904, Amsterdam, 1098 XH, the Netherlands.}
\author{Lex Kaper}
\affiliation{Anton Pannekoek Institute for Astronomy, University of Amsterdam, Science Park 904, Amsterdam, 1098 XH, the Netherlands}
\author{Alex de Koter}
\altaffiliation[Also at ]{Institute of Astronomy, KU Leuven, Celestijnenlaan 200 D, Leuven, 3001, Belgium.}
\affiliation{Anton Pannekoek Institute for Astronomy, University of Amsterdam, Science Park 904, Amsterdam, 1098 XH, the Netherlands}
\altaffiliation[Also at ]{Institute of Astronomy, KU Leuven, Celestijnenlaan 200 D, Leuven, 3001, Belgium.}
\author{Tomer Shenar}
\affiliation{Departamento de Astrof´ısica, Centro de Astrobiolog´ıa (CSIC-INTA),
Ctra. Torrej´on a Ajalvir km 4, 28850 Torrej´on de Ardoz, Spain.}
\author{Steven Rieder}
\altaffiliation[Also at ]{Institute of Astronomy, KU Leuven, Celestijnenlaan 200 D, Leuven, 3001, Belgium.}
\affiliation{Anton Pannekoek Institute for Astronomy, University of Amsterdam, Science Park 904, Amsterdam, 1098 XH, the Netherlands}

\date{\today}

\begin{abstract}
  The $\sim 60\,000$ solar-mass (\MSun) star-cluster R136 (NGC~2070)
  in the Tarantula Nebula in the Large Magellanic Cloud is the host of
  at least 55 massive stars ($M \apgt 10$\,\MSun) which move away from
  the cluster at projected velocities $\gteq 27.5$\,km/s
  \cite{2024Natur.634..809S}.  The origin of the high velocities of
  such runaway stars have been debated since the 1960s, resulting
  either from dynamical ejections
  \citep{1961BAN....15..265B,1961BAN....15..291B} or from supernova
  explosions \citep{1983ApJ...267..322H}.  Due to the Gaia satellite's
  outstanding precision, we can now retrace the most recently ejected
  binary star, Mel 34, back to the center of R136 and reconstruct the
  events that 52\,000 years ago let to its removal from R136, i.e., we
  establish its dynamical interaction and ejection history. We find
  that this ejection requires the participation of 5 stars in a strong
  interaction between a triple composed of the tight massive binary
  Mel~39 orbited by the star VFTS~590, and the binary star Mel~34.
  The participation of 5 stars is unexpected because runaway stars
  were not expected to result from triple interactions
  \cite{2011Sci...334.1380F}.  The deterministic nature of the
  Newtonian dynamics in the scattering enables us to reconstruct the
  encounter that ejected Mel~34. We then predict that Mel~39 is a
  binary star with an 80\,\Msun\, companion star that orbits within
  $\sim 1^\circ$ in the same plane as Mel~34, and escapes the cluster
  with a velocity of $\sim 64$\,km/s.  The five stars will undergo
  supernova explosions in the coming 5\,Myr at a distance of $\sim
  180$\,pc to $\sim 332$\,pc from their birth location (R\,136). The
  resulting black hole binaries, however, are not expected to merge
  within a Hubble time.
\end{abstract}

\maketitle

\renewcommand{\theequation}{\arabic{equation}}
\setcounter{equation}{0}
\renewcommand{\thetable}{\arabic{table}}
\setcounter{table}{0}
\renewcommand{\thefigure}{\arabic{figure}}
\setcounter{figure}{0}
\renewcommand{\thesection}{\arabic{section}}
\renewcommand{\thesubsection}{\thesection.\arabic{subsection}}
\setcounter{section}{0}
\setcounter{secnumdepth}{2}

\section{Introduction}

The young ($\aplt 1$\,Myr) and massive ($\sim 60\,000$\,\MSun) star
cluster NGC2070 (also known as R136) in the Large Magelanic cloud is
one of the most dynamically active regions in the local group
\citep{2010ARA&A..48..431P}.  It is host of the largest known
population of stars more massive than 100\,\MSun\, and several even in
excess of 150\,\MSun\,
\citep{2010MNRAS.408..731C,2022A&A...663A..36B}.

With a core-density of $n \simeq 4.2\times 10^5$ stars/pc$^{3}$, and a
mean stellar mass of $\langle m \rangle \sim 3.6$\,\MSun\, in the
segregated core of R136, stellar encounters are common and violent,
leading to frequent stellar collisions \citep{2007Natur.450..388P} and
high-velocity ejections \citep{2011Sci...334.1380F}.  Stoop et al
(2024)\nocite{2024Natur.634..809S} reported on 20 stars of $\apgt
10$\,\MSun\, escaping the cluster with a velocity of $>27.5$\,km/s in
the last 0.84\,Myr, leading to a mean ejection rate of $\Gamma \apgt
24$\,Myr$^{-1}$.

These runaways cannot have been produced in the supernova explosion in
close binary systems \cite{1961BAN....15..291B,1961BAN....15..265B}
because with an age of 1.5--2\,Myr \cite{2024Natur.634..809S} the
cluster is too young for the first supernova to occur. These observed
runaway stars must then have been violently ejected in multi-body
encounters \cite{1954ApJ...119..625B,Poveda_1967} or in the
kinematically cold collapse in a multi-clump cluster merger
\cite{2024A&A...690A.207P}. The latter process explains about half the
runaways ejected in the recent wave, particularly those in a preferred
direction and with velocities around 40\,km/s, as discussed in
\cite{2024Natur.634..809S,2024A&A...690A.207P}.

The most recently ejected runaway binary Mel~34 is moving too fast to
be ejected through a multi-clump cluster merger, and must have been
ejected in a strong encounter with other stars. Here, we present a
forensic analysis of the last encounter that led to the dynamical
ejection of Mel~34. We reconstruct the details of this encounter and
identify the participating other stars, which were ejected in the same
event.


\section{The last massive runaway star Mel~34}

The most recently ejected object is the binary Mel~34AB composed of
two stars of $139^{+21}_{-18}$\,\MSun\, and $127\pm17$\,\MSun\, in an
eccentric ($e = 0.68\pm 0.02$) $154.55\pm 0.05$\,day ($a \simeq
3.63$\,au) orbit. Mel~34 was ejected $52 \pm 8$\,kyr ago with a
projected velocity of $\sim 46.3\pm6.4$\,km/s
\cite{2019MNRAS.484.2692T}.

\begin{figure}
\center
\includegraphics[width=1.0\textwidth]{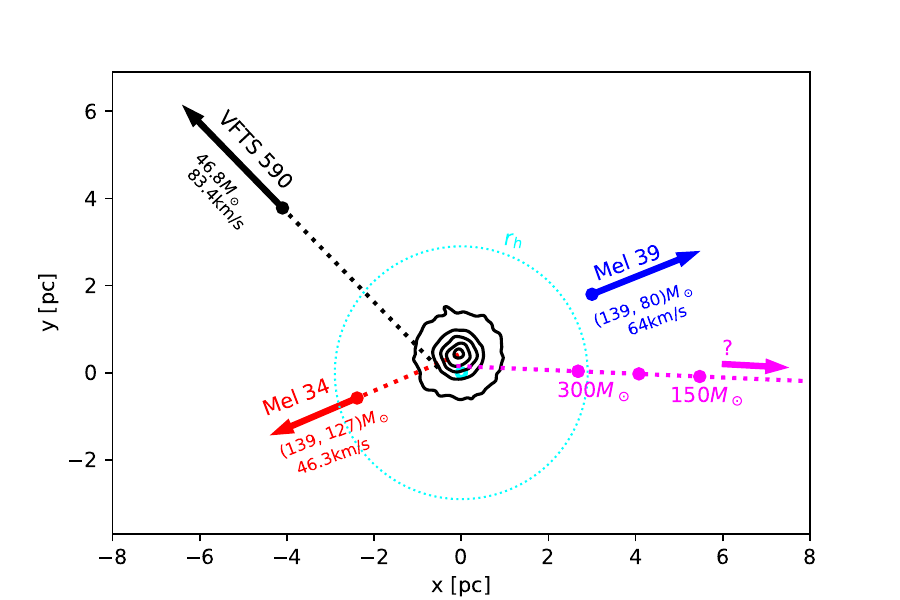}
\caption[]{Reconstruction of the historic trajectory for the massive
  binary Mel~34 (red), VFTS~590 (black) and the mystery runaway star
  that would conserve angular momentum in a binary-binary encounter
  (magenta). The arrows are proportional to the velocities (except the
  magenta one, which is mass-dependent). The cyan-colored circles
  indicate the cluster's core and half-mass radii. The contours give
  the probability density distribution for the encounter site at 0.01,
  0.2, 0.4, etc.  confidence levels. The blue dot to the right
  indicates where Mel~39 is observed, the arrow shows our predicted
  velocity vector.
\label{fig:Mel34_and_VFTS590}
}
\end{figure}

The characteristics of Mel~34 are surprisingly consistent with a
dynamically formed binary \citep{2000IJoMP...15..4871P}: the system
exhibits a tight orbit with high eccentricity, and both stars are
among the most massive in the cluster.  In addition, tracing back
Mel~34's trajectory, it passed 0.046 to 0.060\,Myr ago to within the
projected position of $(-0.058 \pm 0.65, 0.42 \pm 0.28)$\,pc from the
estimated center of R136 \cite{2024Natur.634..809S}, as depicted
in~\cref{fig:Mel34_and_VFTS590}.

\section{The ballistics of the last runaways}

Ejecting a massive binary with such a high velocity requires an
unusually strong encounter with other stars. The binary itself can
deliver enough energy for such an ejection if the encountering star is
of comparable mass, but can we identify this other star?

Mel 34 may have formed as a wide binary system, over time decreasing
its orbital separation \cite[a process called
  hardening,][]{1975MNRAS.173..729H} through interactions with its
cluster siblings and ambient gas. At orbital separations less than $a
\sim 1400$\,au its binding energy would exceed $E_{\rm bin} \apgt
10$\,kT (see~appendix\,\cref{SM:binaries}, on the use of kT as a unit of
energy). At a $\sim 60$\,au ($\sim 600$\,kT) interactions become
sufficiently energetic for ejecting other stars. From that moment,
such a binary is refered to as a bully binary. These bully binaries
play a prominent role in the dynamical evolution of stellar clusters
\citep{2011Sci...334.1380F}.  Although not currently a cluster member,
Mel~34 today has an equivalent binding energy of $E_{\rm bin} \sim
4300$\,kT, making it a bully binary.

The typical energy release in an encounter between a target bully
binary (with mass $m_t$) and a projectile star with mass $m_p$ is
\cite{2003gmbp.book.....H}
\begin{equation}
dE_{\rm bin} = \xi q E_{\rm bin}.
\label{Eq:dEbin}\end{equation}
Here $q \equiv m_p/m_t$, and $\xi = {\cal O}(0.2)$ (a dimension-less
scaling factor depending on equipartition \citep[see ][eqs. 4.29 and
  5.40]{1975MNRAS.173..729H}, although the precise value depends on
the details of the encounter).  A bully binary leads to roughly 8
escapers of comparable mass before it is sufficiently hard to eject
itself \cite{1992ApJ...389..527H}.  When Mel~34 ejected itself in a
dynamical interaction in R136's core, the encountering object being
lower in mass was also ejected because linear momentum in such
interactions is conserved.


The $46.8^{+6.5}_{-6.1}$\,\MSun\, star VFTS~590 was ejected
$62^{+5}_{-4}$\,kyr ago from $(-0.47, 0.31)$\,pc relative to the
cluster center with a velocity of $83.4\pm6.2$\,km/s ($\sim 324$\,kT,
see also figure\,\ref{fig:Mel34_and_VFTS590})
\cite{2024Natur.634..809S}.  Both stars, Mel~34 and VFTS~590, are then
ejected at the same time within $1.1$ standard deviations.  VFTS~590
was launched in a direction almost perpendicular to the bulk of the
other recently ejected stars, suggesting that it does not belong to
the population of runaways accelerated in the kinematically cold
sub-cluster merger.  However, the amount of kinetic energy required to
eject Mel~34 and VFTS~590 exceeds $\sim 891$\,kT; more than 8 times
the expected $dE_{\rm bin} \sim 107$\,kT generated in such an
encounter (see~\cref{Eq:dEbin}).

The alternative exchange interaction in which, for example, VFTS~590
is the original secondary in a binary with Mel\,34B, can deliver the
appropriate amount of energy, but to account for the kinetic energy of
the two runaways and Mel~34's binding energy, the pre-encounter
orbital separation of the binary should be between $0.8$\,au and
$3$\,au. Such an interaction typically has a 17 times higher
probability to result in a collision between two or more stars rather
than a clean exchange. The latter cross section is only $\sigma \sim
11$\,au$^2$, or equivalently expressed in the encounter rate $\Gamma
\sim 83$\,Myr$^{-1}$, and the typical velocity with which VFTS~590
would be ejected is much too high (up to $\sim 140$\,km/s).  VFTS~590,
however, cannot have been responsible for the ejection of Mel~34
because the interaction does not conserve angular momentum
(see~\cref{SM:momentumconservation}).

\section{The other stars participating in the encounter}

Angular momentum can only be conserved if at least one other star
participated in the encounter, which would also have been ejected from
the cluster (\cref{fig:Mel34_and_VFTS590}). We performed 4-body
(binary-binary, and triple-single) scattering experiments to determine
the most favorable mass of the unknown runaway
(see\,\cref{sect:crosssections}). The largest cross-section and
consistent post-encounter parameters are achieved for a
$90\pm3$\,\MSun\, star: which would be VFTS~590's binary companion
before the encounter (see~\cref{SM:Mel39}). This mystery runaway star
would now be at a distance of about $8.5\pm0.3$\,pc from R136's
center. In~\cref{fig:Mel34_and_VFTS590}, we show the escaping star's
trajectory that conserves linear momentum. There is no prominent
candidate star with a measured proper motion along this trajectory up
to a distance of $\sim 600$\,pc \cite{2024Natur.634..809S}, although
it could be obscured, or below the detection limit (of $\sim
10$\,\MSun\, for a main-sequence star).  We consider the possibility
that Mel~34 was ejection through the interation with an
intermediate-mass black hole exotic because the cluster is
insufficienly dense and too young to have formed such an object
\cite{2002ApJ...576..899P}.


Almost directly opposite Mel~34 at the other side of the core of
R~136, at an angle of $\sim 179.8^\circ$, another binary called Mel~39
runs away from R\,136. The Wolf-Rayet primary star in Mel~39 is $\sim
140$\,\MSun. An orbital period of about 92\,days is reported, but the
companion star is not yet identified \cite{2020MNRAS.499.1918B}.

Mel~39 is about 20\% further away (in projection) from the center of
R~136 than Mel~34 (see~\cref{fig:runaway_mass_function}). If both
objects ejected each other from the cluster center, the projected
velocity for Mel~39 would be about 64\,km/s. With the conservation of
linear momentum, we estimate the total mass of Mel~39 $\sim
220$\,\MSun; that leaves about 80\,\MSun\, for its companion star.
Interestingly, this is consistent with the spectroscopically
determined mass of $80\pm 11$\,\MSun, derived by
\cite{2025MNRAS.539.1291P}.  It is conceivable that such a secondary
was not observed in optical surveys because the primary apparently
outshines it. With these parameters, its orbital separation will be
about 2.48\,au, somewhat tighter than Mel~34, but with a binding
energy of $\sim 4000$\,kT, it is softer.

A single star (VFTS~590) that simultaneously engaged two binaries in a
strong encounter is improbable and leads to too high runaway speeds
except if the relative encounter-velocity $\apgt 15$\,km/s
(see~\cref{SM:R136}). The latter exceeds the cluster's escape speed,
making such an encounter highly unlikely. The star VFTS~590 then must
have orbited one of the interacting binaries.

\section{Reconstructing the interaction}

Exploring the entire parameter space of a 5-body encounter is hindered
by its large volume. We can limit the available parameter space
enormously by requirements that energy, and momentum (linear and
angular) are conserved througout the interaction. 

The two binaries, currently with binding energies of $\sim 4300$\,kT
and $\sim 4000$\,kT for Mel~34 and Mel~39, respectively, have to
generate enough energy to accelerate VFTS~590 (324\,kT), Mel~34
(567\,kT), and Mel~39 (896\,kT) to their present-day velocities.  This
total of 1787\,kT has to come from the binding energies of the binary
and the triple.  Note that for an equal-mass encounter, the delivery
of 1787\,kT can satisfactorily be achieved by two binaries with a
total binding energy of $\sim 8935$\,kT, which is only slightly higher
than the total binding energy in today's two binaries, leaving
roughtly 635\,kT for the orbit of the tertiary. We then argue that the
binding energies of the two binaries is rather well constrainted, and
we only have to identify the various components.  Note that the
minimal stable orbit for VFTS~590 as tertiary component around Mel~34
or Mel~39 has a semi-major axis of $\sim 25.1$\,au or 36.6\,au,
respectively \cite{2001MNRAS.321..398M}.  The binding energy of these
orbits corresponds to 300\,kT and 362\,kT, respectively.

To simplify the discussion, we parametrize the interaction as a target
triple $t$ with inner components \tAa\, and \tAb, and outer component
\tB. The projectile binary comprises two stars \pA\, and \pB. We then
write the pre-encounter topology as $\langle$((\tAa, \tAb), \tB),
(\pA,\pB)$\rangle$. Here, the angle brackets indicate a hard dynamical
encounter, and parenthesis the bound pairs.

\begin{table}
\begin{tabular}{llrrrr}
\hline
id & initial configuration & $\sigma$ & $f_{\rm esc, VFTS590}$& $f_{\rm run}$ & $\Gamma_{\rm esc, VFTS590}$ \\
& & [$10^6$ au$^2$] & & & Myr$^{-1}$ \\
\hline
A & $\langle$((\tAa, \tAb), \tB), (Mel39B, VFTS~590)$\rangle$ & 61.8 & 0.002 & 0.000 & 0.0 \\
B & $\langle$((\tAa, \tAb), VFTS~590), (\pA, Mel39B)$\rangle$ & 53.5 & 0.397 & 0.016 & 6.2 \\
C & $\langle$((\tAa, \tAb), Mel39B), (\pA, VFTS~590)$\rangle$ & 38.6 & 0.002 & 0.000 & 0.0 \\
D & $\langle$((\tAa, VFTS~590), \tB), (\pA, Mel39B)$\rangle$ & 90.5 & 0.129 & 0.002 & 1.1 \\
E & $\langle$((\tAa, Mel39B), \tB), (\pA, VFTS~590)$\rangle$ & 63.0 & 0.004 & 0.000 & 0.0 \\
F & $\langle$((\tAa, VFTS~590), Mel39B), (\pA, \pB)$\rangle$ & 106 & 0.099 & 0.007 & 5.2 \\
G & $\langle$((\tAa, Mel39B), VFTS~590), (\pA, \pB)$\rangle$ & 101 & 0.442 & 0.049 &34.9 \\
H & $\langle$((Mel39B, VFTS~590), \tB), (\pA, \pB)$\rangle$ & 109 & 0.109 & 0.006 & 4.6 \\
\hline
\end{tabular}
\caption[]{The 8 initial configurations for a triple-binary encounter.
  The target triple has an inner separation $a_{\rm in} = 4$\,au, and
  outer separation $a_{\rm out} = 40$\,au.  The projectile binary with
  $a_{\rm p} = 6$\,au, encounters the terget triple with a relative
  velocity of $v_{\rm enc} = 7$\,km/s. The last four columns give the
  interaction cross-section ($\sigma$), the fraction of cases in which
  VFTS~590 is ejected ($f_{\rm esc, VFTS590}$), with a velocity of at
  least $v_{\rm ej} \apgt 27.6$\,km/s ($f_{\rm run}$), and the
  corresponding rate that VFTS~590 and the two surviving binaries are
  ejected with at least $v_{\rm ej} \apgt 27.6$\,km/s calculated for
  the core collapsed cluster R136 ($\Gamma_{\rm esc, VFTS590}$). The
  most likely scenario is G.}
\label{tab:cross_sections}
\end{table}

The various stars can be divided into a triple and a binary in $5!$
combinations.  We reduce parameter space by adopting the masses of
VFTS~590 ($50$\,\MSun) and Mel~39B ($80$\,\MSun) unique, and assuming
$130$\,\MSun\, for Mel~34A, Mel~34B, and Mel~39A. We further adopted
circular orbits for the projectile binary and the inner binary of the
target triple.  Relaxing this assumption does not affect the
energetics of the encounter.  Although, we cannot exclude that Mel~34
has some residual eccentrity from a previous encounter, the tidal
circularization time scale for Mel~39 is smaller than its lifetime
\citep[see Eq.\,4.13 of][]{1977A&A....57..383Z}, and it was probably
circularized before the encounter.  The eccentricity of the tertiary
was taken randomly from the thermal distribution with the requirement
of a stable triple (see\,\cref{tab:cross_sections}).

We then constrain the pre-encounter parameters by determining the most
probable configuration, and in addition to that, identify which of the
5 component should be associated with VFTS~590 or Mel~39B. The
resulting 8 combinations are listed in table\,\ref{tab:cross_sections}
as scenario A till H (see also \cref{sect:crosssections}).

The optimal parameters are determined by running 5-body simulations to
reproduce the post-encounter binaries, Mel~34 and Mel~39, with their
appropriate orbital separations of 3.63\,au and 2.48\,au, respectively
and runaway speeds (see\,\cref{sect:crosssections}). With an
interaction rate of $34.9$\,Myr$^{-1}$, we favor scenario G. VFTS~590
then identifies with \tB, the tertiary orbiting a binary, with Mel~39B
as the secondary (\tAb). The presented calculations assume an orbital
separation of 4\,au for the inner (target) binary of the triple (\tAa,
\tAb), 40\,au (up to 200\,au) for the outer orbit, and 6\,au for the
projectile binary (\pA, \pB).

To further identify the three remaining stars, \tAa, \pA and \pB, we
calculate the relative frequency for each of the various
configurations for the interactions in scenarios G
(\cref{tab:interaction_frequency} in \cref{sect:crosssections} lists
all relative probabilities for scenarios A till H).

The binary-preserving encounters for model G have the largest
probability with a rate of $\sim 27$\,Myr$^{-1}$; conveniently close
to the observed $24$\,Myr$^{-1}$. The most probable interaction then
is {\small
\begin{eqnarray}
\langle (({\rm Mel}39A, {\rm Mel}39B), {\rm VFTS}590), ({\rm Mel}34A, {\rm Mel}34B) \rangle \rightarrow
 ({\rm Mel}34A, {\rm Mel}34B), ({\rm Mel}39A, {\rm Mel}39B), {\rm VFTS}590
\end{eqnarray}
}
The interaction
{\small
\begin{equation}
\langle (({\rm Mel}34A, {\rm Mel}39A), {\rm VFTS}590), ({\rm Mel}34B, {\rm Mel}39B) \rangle \rightarrow \ldots
\end{equation}
}
leading to the same final configuration is less probably by a factor
$\sim 5$ ($\Gamma\simeq 6$\,Myr$^{-1}$).

\begin{figure}
\center
\includegraphics[width=1.0\textwidth]{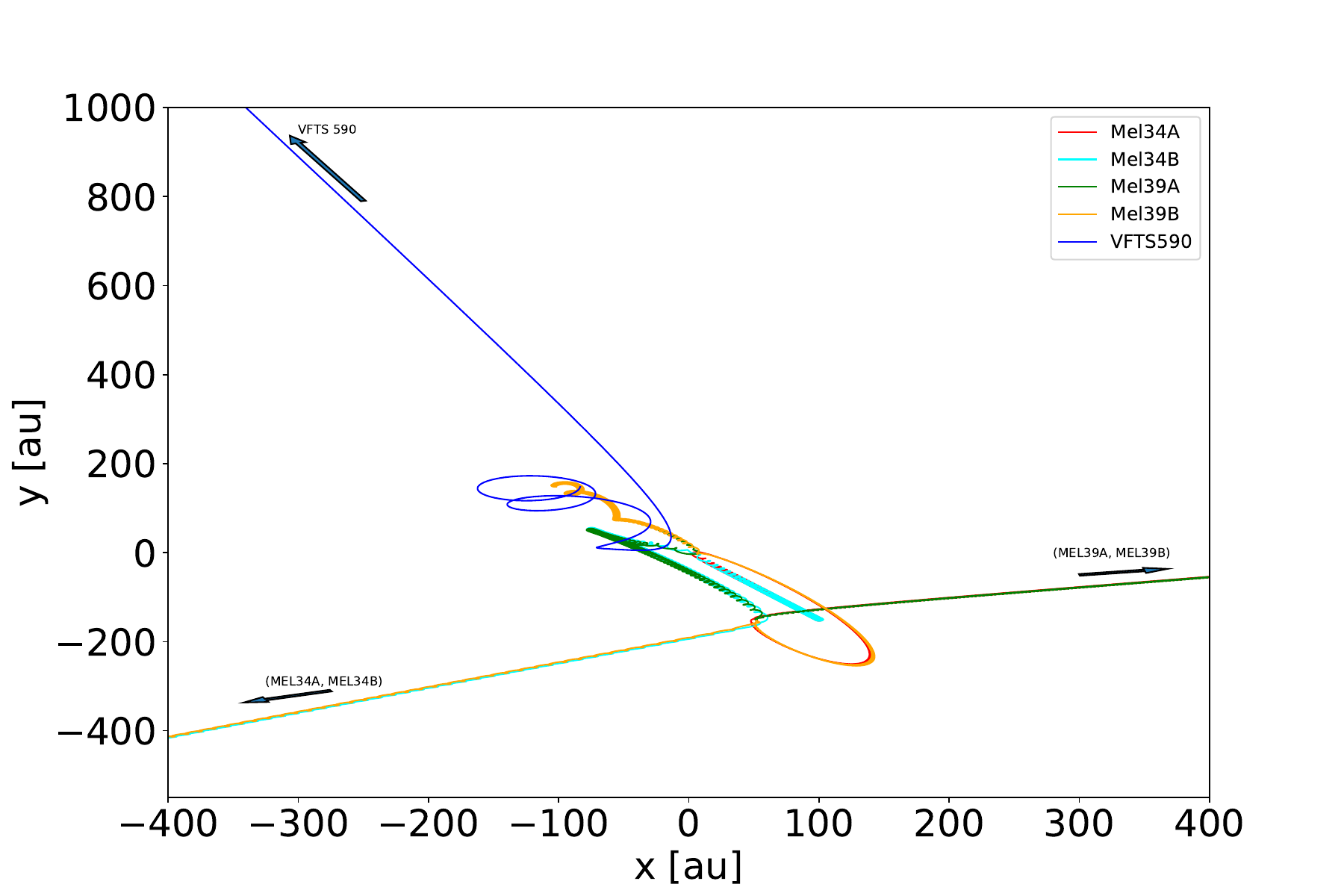}
\caption[]{ Interaction between the triple ((Mel~34A, Mel~39B),
VFTS~590) and the binary (Mel39A, Mel34B) for scenario B in a strong
dynamical interaction with a relative velocity of 7\,km/s. The
interaction leads to the ejection of VFTS~590 as a single star, and
the preservation of the two binaries. The two escaping binaries are
ejected in opposite directions, consistent with the observations,
including the ejection velocities, the relative orbital separations
and the eccentricity of Mel~34.
\label{fig:Mel34_VFTS590_and_Mel39_interacting}
}
\end{figure}

Figure \,\ref{fig:Mel34_VFTS590_and_Mel39_interacting} presents an
example encounter showing interaction G. The cross section is largest
for orbital separations of the inner target and projectile binaries of
4\,au and 6\,au, respectively, and these reproduce the current
observed orbital separations and runaway velocities. The orbital
separation of VFTS~590 around Mel~39 (in the range of 40\,au to
200\,au) has a negligible effect on the interaction cross-section
because the energetics of the interaction is dominated by the binding
energies of the two binaries (see~\cref{sect:crosssections}).

Changing the incoming relative encounter velocity affects the result
as expected through gravitational focusing. We prefer a relative
encounter velocity of 7\,km/s because it is consistent with the
cluster's core velocity dispersion and the resulting cross-section is
consistent with the estimated encounter rate for the cluster center
(see~\cref{sect:crosssections}).


\section{Results}

The last energetic encounter in R\,136 occurred $\sim 52\,000$ years
ago between a hierarchical triple and a binary. The triple was
composed of ((Mel~39A, Mel~39B), VFTS~590) with orbital separations of
$4.0\pm0.5$\,au for the inner target orbit, and an outer orbit between
$40$\,au and $200$\,au. Before the encounter, the binary (Mel~34A,
Mel~34B) had an orbital separation of $6.0\pm0.5$\,au.

\begin{figure}
\center
\includegraphics[width=1.0\textwidth]{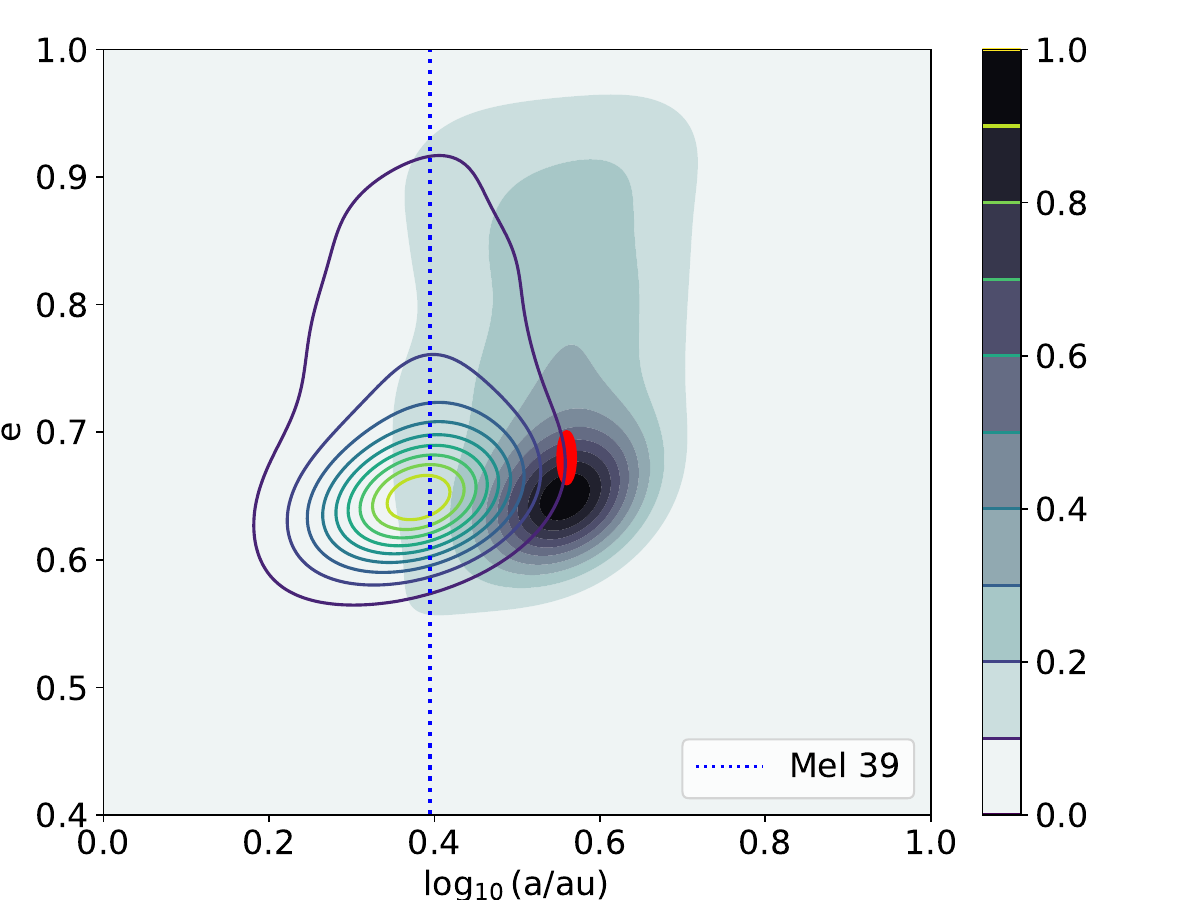}
\caption[]{Orbital separation vs eccentricity for Mel34 (gray shades,
  see color-bar to the right) and Mel39 (contours, overplotted in the
  colorbar). The two distributions are generated from the preserving
  interaction with an initial separation of 4\,au for Mel39, 40\,au
  for VFTS~590 around Mel~39, and 6\,au for Mel~34. The red ellipsoid
  point gives the observed orbital elements for Mel~34 (its size is 2
  standard deviations), and the vertical blue dotted line gives the
  observed semi-major axis (assuming an 80\,\MSun\, star for the
  secondary) for Mel~39.  }
\label{fig:sma_vs_ecc_Mel34_Mel39}
\end{figure}

The interaction reproduces statistically the observed orbital
separations of Mel~34 and Mel~39, the velocities of both binaries, and
the observed velocity of VFTS~590. Table \,\ref{tab:final_orbits}
presents the pre- and post-encounter parameters. In
figure\,\ref{fig:sma_vs_ecc_Mel34_Mel39} we present the distribution
of semi-major axis and eccentricity for the post-encounter binaries
Mel~34 (shades) and Mel~39 (contours).  The orbital separations for
both binaries are well constrained by the simulations and match the
observed values. The relatively small dispersion in probability
density distribution for the eccentricities comes as a surprise, as we
naively would have expected a close-to-thermal distribution.

In our best solution Mel~39A is accompanied by an $80$\,\MSun\, star
in a $2.48$\,au orbit. The interaction between the two binaries
induced a high eccentricity ($e = 0.78\pm0.09$ but with a median of
$0.73\pm0.02$) in Mel~39, and a mean runaway velocity $v_{\rm run} =
70.6\pm8.6$ (median is $60.7$\,km/s). The two binaries are ejected in
almost opposite directions with a median angle of $\Theta \simeq
172^\circ$ ($\langle \Theta \rangle = (152.9\pm24.2)^\circ$). The
relative inclination between Mel~34 and Mel~39 is $\delta i = (-1.1
\pm 56)^\circ$. Our derived inclination for Mel~39 is consisent with
the recent observed value of $90^\circ$ \cite{2025MNRAS.539.1291P} and
comparable to their observed eccentricity of $0.618\pm 0.014$.  But
\cite{2025MNRAS.539.1291P} arrive at a somewhat wider orbit.  The
argument of pericenter and the line of the ascending node do not show
any systematic correlations in the simulation results.

\begin{table}
\begin{tabular}{lrrrrrr}
\hline
Binary & & Tertiary & $a_{\rm init}$& $a_{\rm fin}$ & $e_{\rm fin}$ & $v_{\rm run}$ \\
& & & [au] & [au]& & km/s \\
\hline
((Mel39A &Mel39B)& &$4$& $1.39\pm 0.57$ & $0.84\pm0.10$ & $58.7 \pm3.8$\\
& & VFTS~590) &$40-200$& & & $65.1\pm 9.2$\\
(Mel34A & Mel34B)& &$6$& $6.03 \pm 9.14$ & $0.74 \pm 0.08$ & $70.6 \pm 8.6$\\
\hline
\end{tabular}
\caption[]{Pre and post-encounter orbital parameters for the two
  binaries and the tertiary component. The first row identifies the
  binary components followed by the name of the third star.  The
  subsequent column gives the initial orbital separation ($a_{\rm
    init}$). The last three columns give the final semi-major axis
  ($a_{\rm fin}$), eccentricity ($e_{\rm fin}$) and the mean ejection
  velocity (with respect to the barycenter, $v_{\rm run}$).}
\label{tab:final_orbits}
\end{table}

\section{Discussion and conclusions}

The two bully binaries can deliver sufficient energy to eject all
the 20 observed runaways of $\apgt 10$\,\MSun\, in the last 0.84\,Myr.
The asymmetry in the direction of the recent runaways
\cite{2024Natur.634..809S}, however, suggests that about half of them was
ejected in a cold sub-cluster merger, as discussed in
\cite{2024A&A...690A.207P}. Considering the top-heavy mass function and the
hardness of the two bullies ($2470$\,kT for Mel~39, and $2230$\,kT for
Mel~34), we then expect that $\sim 52$ more runaways will be
discovered with a mass between 1\,\MSun\, and the observational
detection limit of $\sim 10$\,\MSun; with early ejected runaways having
systematically lower velocities.

Both bully binaries ran into each other about 52\,000 years ago,
leading to their mutual ejection; this epoch of bully binaries has now
temporarily ceased. Mel~34 and Mel~39 may have prevented the early
hardening (from $\sim 10$\,kT to $1000$\,kT) of a new bully, and it
may take a while before another binary takes their place. The four
binaries R136~38, R136~39 (not to be confused with Mel~39), R136~42,
and R136~77 \cite{2002ApJ...565..982M} have such short orbital periods
that a strong interaction is expected to lead to a collision rather
than a dynamical ejection.

In about $3$\,Myr, both binaries Mel~34 and Mel~39 will circularize,
experience mass transfer, and eventually, after about $5$\,Myr, the
stars explode in supernovae. By that time Mel~34 as traveled $\sim
184$\,pc, and Mel~39 $\sim 258$\,pc. Mel~34 and Mel~39 are the
prototypical example binaries that eventually lead to black-hole
mergers \cite{2018LRR....21....3A}, though not in a Hubble time.

We demonstrated how observing the physical properties of runaway stars
provides the opportunity to reconstruct in detail the dynamical
history of young-massive stellar clusters.

\section*{Data and code availability}

All data is available at zenodo
\url{https://zenodo.org/uploads/10256841} under DOI:
10.5281/zenodo.10256841, and all the source code is on github
\url{https://github.com/amusecode/Starlab}.





\begin{appendix}

\renewcommand{\theequation}{\thesection.\Roman{equation}}
\setcounter{equation}{0}
\renewcommand{\thetable}{\thesection.\Roman{table}}
\setcounter{table}{0}
\renewcommand{\thefigure}{\thesection.\Roman{figure}}
\setcounter{figure}{0}
\renewcommand{\thesection}{\Alph{section}}
\renewcommand{\thesubsection}{\thesection.\arabic{subsection}}
\setcounter{section}{0}
\setcounter{secnumdepth}{2}

\makeatletter
\renewcommand{\c@secnumdepth}{0}
\makeatother

\noindent
{\large {\bf Appendix}}
\bigskip

To reconstruct the last encounter that ejected Mel~34 from the star
cluster NGC2070 (R136), we performed more than 2 million scattering
experiments for 3, 4 and 5 bodies, 12 full N-body simulations of the
entire star cluster, 5 single stellar-evolution calculations, and 2
binary-evolution calculations. In all simulations, we adopt the
zero-age main-sequence radii of the stars calculated using the AMUSE
{stellar\_simple.py} script \cite{10.1088/978-0-7503-1320-9}. The
radii for Mel34A, Mel34B, and Mel\,39A are then $\sim 17$\,\RSun\,
\cite{2019MNRAS.484.2692T}. The two stars Mel\,39B and VFTS\,590 are
15\,\RSun\, and 11\,\RSun, respectively.

\section{The cluster R136}\label{SM:R136}

The stars in a cluster are usually distributed with a power-law
mass-function with a slope of $-2.35$ between the hydrogen-burning
limit (of about 0.08\,\MSun) and some maximum mass (of $\sim
100$\,\MSun). This typical mass function has a mean mass of $\sim
0.35$\,\MSun\, \cite{1955ApJ...121..161S}, but the core population of
a collapsed cluster is considerably more massive. We numerically
explore the core-mass function of a 60\,000\,\MSun\, non-mass
segregated and relatively cold ($Q=0.1$ to virializd $Q \equiv E_{\rm
kin}/E_{\rm pot}=0.5$) Plummer \citep{1911MNRAS..71..460P} sphere
with a half-mass radius of 2.9\,pc. Most simulated clusters virialize
within $\sim 0.2$\,Myr, at which point, the core has shrunk to $\sim
0.10$\,pc \cite{2014MNRAS.445..674C}. By this time, the mean mass
within 0.06\,pc is about 11\,\MSun\, (with an average of $\sim
3.6$\,\MSun\, in the core). At this time the density in the cluster
core $n \simeq 4.2\times 10^5$ stars/pc$^{3}$, and $\rho \simeq
1.5\times 10^{6}$\,\MSun/pc$^3$.

\begin{figure}
\center
\includegraphics[width=1.0\textwidth]{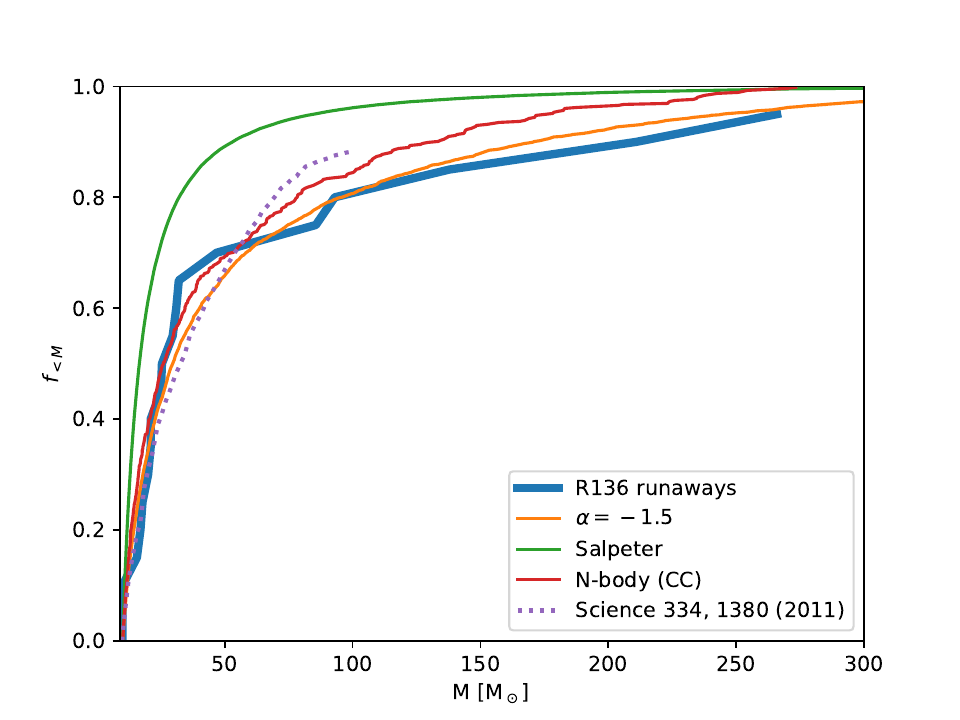}
\caption[]{cumulative distribution of the recent ($t_{\rm ej}<1$\,Myr)
runaway stars (blue \cite{2024Natur.634..809S}). The mass function of
the core population of 5 core collapsed star-cluster simulations is
presented in red, which is slightly steeper (power-law slope $-1.7$)
than the runaway mass function, but the measurement was taken just
before the first bully binary formed. The mass function within $0.8
r_{\rm core}$ is indistinguishable from the runaway mass
function. Overplotted (in orange) is a power-law with a slope of
$-1.5$, the Salpeter (green) mass function (power-law with slope
$-2.35$).
\label{fig:runaway_mass_function}
}
\end{figure}

The typical ejected star will follow the top-heavy core mass function
\cite{2011Sci...334.1380F}. In
figure\,\ref{fig:runaway_mass_function}, we compare the top-heavy mass
function of the recent runaways with the Salpeter slope.

With a core radius of $r_{\rm core} \simeq 0.1$\,pc, and a half-mass
radius of 2.9\,pc, the cluster NGC2070 (or R136) resembles a King
profile with a dimensionless depth of the central potential of $W_0
\sim 9.8$ \cite{1966AJ.....71...64K}. Such a King model has a relative
core mass of $0.017$, or $\sim 10^3$\,\MSun\, and a velocity
dispersion of $\sim 7$\,km/s. As a consequence, R136 is in a state of
gravothermal collapse \cite{1991ApJ...383..181M}, and prone to strong
dynamical interactions.

\section{Soft and Hard binaries}\label{SM:binaries}

To mediate the discussion, it helps to consider the cluster as a gas
and define the unit of energy in terms of kT $\propto mv^2$ as the
kinetic encounter energy between stars in the cluster core
\cite{2003gmbp.book.....H}. With a mean stellar mass in the cluster
core of 3.9\,\MSun\, to 11.0\,\MSun\, (see\,\cref{SM:R136}), the
core's kinematic unit for energy $1$\,kT$\simeq 1.1\cdot
10^{46}$\,erg. For convenience we adopt as definition for $E_{\rm kT}
\equiv 10^{46}$\,erg.

The binding energy of a binary with orbital separation $a$ and two
stars of masses $M$ and $m$ is
\begin{equation}
E_{\rm bin} = {GMm \over 2a}
\end{equation}

A binary is generally considered hard once it's binding energy exceeds
10kT. Mel~34, residing in the cluster core would be hard at an
orbital separation of about 1400\,au.

Ejecting a 10\,\MSun\, star from R\,136 with a velocity of 27.5\,km/s
would require a total kinetic energy of $\sim 6.8$\,kT. Mel~34 can
generate this much energy in an encounter if it has a binding energy
of $E_{\rm bin} \apgt 910$\,kT (see \cref{Eq:dEbin} in the main
paper), or an orbital separation of $\sim 15.5$\,au. From 15.5\,au
down to the observed $3.63$\,au Mel~34 has liberated about 3000\,kT in
kinetic energy.  This is enough to eject 439 stars of 10\,\MSun, or 3
stars of 100\,\MSun\, at 100\,km/s. Bear in mind, though,that every
time Mel~34 kicks out a star, it also propels itself in the opposite
direction.

\section{The role of VFTS590}\label{SM:momentumconservation}

The encounter that led to the ejection of Mel~34 must conserve linear
and angular momentum. If Mel~34 ejected VFTS~590, another object must
be ejected along the magenta line in \cref{fig:Mel34_and_VFTS590}
(main paper).  There is no prominent candidate star with a measured
proper motion along this line up to a distance of $\sim 600$\,pc
\cite{2024Natur.634..809S}.

If we assume that a 15\,\MSun\, star, barely observable at the
distance of the LMC, is the star we are looking for; it then was ejected
with a velocity of $\sim 990$\,km/s or a kinetic energy of $\sim
46500$\,kT to conserve linear momentum. The binary Mel~34 cannot have
achieved this because if another binary participated in this encounter
(of which this 15\,\MSun\, runaway was a member), it's orbital
separation would have been smaller than the main-sequence radii of
both stars. An encounter can hardly produce such a high velocity but
rather result in a collision \cite{2009MNRAS.396..570G}.

\section{The mystery star}\label{SM:Mel39}

We performed 4-body (binary-binary) scattering experiments using
starlab \cite{2004MNRAS.351..473P} to determine the most favorable
mass of the unknown runaway. An encounter with both binaries at
orbital separations of 2 to 4\,au would suffice to eject all objects
in their appropriate directions with the observed velocity. It
requires the encountering binary, composed of Mel~34A (or Mel~34B) and
VFTS~590 in a 2 to 4 \,au orbit to be broken up, and the primary to
exchange for our unknown runaway star. The largest cross-section and
the most probably post-encounter parameters are achieved for a $\sim
90$\,\MSun\, star. Currently this 90\,\MSun\, mystery runaway star
would be at a distance of about 8.5\,pc from the R136's center moving
away along the magenta line in figure\,\ref{fig:Mel34_and_VFTS590}
with a velocity of about 180\,km/s; such a star is not observed.

We cannot fully exclude the hypothesis that the VFTS~590 and Mel34AB
were ejected through a 4-body encounter in which one of the stars has
a mass below the detection limit of 10\,\MSun.  In that case the most
probable encounter is between a target binary composed of Mel\,34A and
VFTS~590 in a 2 to 4 au orbit that interacts with the projectile
binary composed of Mel\,34B and a $<10$\,\MSun\, star (below the
detection limit). The projectile binary then must have an orbital
separation $\aplt 0.2$\,au, in order to accommodate the post-encounter
energetics. Note that the cross section for this interaction is almost
identical if we swap Mel\,34A for Mel\,34B.

Such an interaction leads to a runaway velocity for VFTS~590 of
$114\pm71$\,km/s with a median of $\sim 99$km/s, consistent with the
observed velocity of $83.4\pm6.2$\,km/s.  The post-encounter binary
Mel\,34AB, would then acquire a median velocity of $\sim 199$\,km/s,
much higher than the observed $64$\,km/s.  Note that these velocities
are acquired under the assumption that all four stars are
point-masses: and no collisions can occur.  The probability, however,
that this interaction leads to a collision between two (or even three)
stars is more than 60 times larger than the clean exchange and
ionization interaction in which VFTS~590 and Mel~34AB would be
ejected. A collision between two or even three of the participating
stars will not lead to high-velocity runaways.  The data for these
calculations are available at Zenodo
(\url{https://zenodo.org/uploads/10256841}).

We do expect that several collisions have occurred in the central
portion of R\,136, but not in the recent encounter that launched
Mel\,34AB.

\section{Calculating the interaction cross sections}\label{sect:crosssections}

\subsection{Automatic cross section detection}\label{sect:sigma3}

{\sc Starlab} has a handy package for the automatic determination of
cross sections, called {\sc sigma3}. We expanded {\sc sigma3} to
calculate multi-body cross sections \cite{2004MNRAS.350..615G}. In
{\sc sigma3}, one starts by launching a projectile single star onto a
target binary with semi-major axis $a=1$ (in dimensionless N-body
units \cite{1986LNP...267..233H}) in the X-Y plane with the periapsis
along the positive X-axis. The center of mass of the multi-body system
was centred on the origin. The single incoming star is initialized at
$100a$ from a random direction and an impact parameter that brings it
within $a$ of the binary. This process is repeated until the desired
surface density of encounters is reached. In our calculations, we
adopted a trial density of $10^3$. Suppose one or more encounters lead
to anything else than a non-resonant preservating interaction. In that
case, the same procedure is repeated, but now the closest approach
distance is between $a$ and a larger distance such that the surface
area equals the binary surface area. Again $10^3$ scattering
experiments are performed, and if any of the experiments result in a
non-preserving encounter this procedure is repeated until all
encounters in the outermost ring result in a preserving encounter.

The cross-section $\sigma$ is subsequently calculated from the largest
impact parameter $b$ for which an encounter leads to any outcome other
than a preserving encounter: $\sigma = \pi b^2 \times n_{\rm
  hit}/N_{\rm exp}$.  Here $N_{\rm exp}$ is the number of scattering
experiments with impact parameter $<b$, and $n_{\rm hit}$ is the
number of cases where the experiment resulted in the desired final
configuration.

\subsection{3-body scattering}

We use {\sc sigma3} to calculate the 3-body interaction cross-section
between a binary and a single star. The three stars, Mel~34A, Mel34B,
and VFTS~590, appear in avarious configurations in the target binary
and projectile incoming star.

We performed a total of 382104 3-body scattering experiments in order
to determine the cross-section for accelerating Mel~34 and VFTS~590 to
a minimum velocity of 27.6\,km/s. We adopted orbital separations for
Mel~34A and Mel~34B between 1\,au and 1500\,au (in 10 equally
separated logarithmic bins), and varied the incoming velocity from
7\,km/s to 10\,km/s with 1\,km/s increments. We conclude that it
requires considerable fine-tuning to have VFTS~590 acquire its
observed velocity, and preserved angular momentum; the cross-section
for this interaction is negligible small.

\subsection{4-body scattering}\label{Sect:Four_body_scattering}

In four-body scattering experiments, we adopt that either two binaries
interact or a triple interacts with a single star. One is considered
the target, and the other is the projectile. The cross sections are
determined automatically using the procedure set out for 3-body
encounters, except that now the projectile may be a binary star with
semi-major axis $a_p$. The package for the 4-body encounter
cross-section determination is called {\sc sigma}, and generalizes
{\sc sigma3} \cite{2004MNRAS.350..615G}.

A total of 344288 4-body scattering experiments were conducted for
this study. These runs were used to further constrain the most
suitable parameters in the case that VFTS~590 would have been a binary
companion before encountering Mel~34. The semi-major axis of the
target binary was varied between 2\,au and 8\,au (with constant
increments of 1\,au), and the projectile binary between 2\,au and
18\,au (with constant increments of 1\,au). The relative encounter
velocity was varied between 3\,km/s and 18\,km/s (with constant
increments of 1\,km/s).

From these calculations, we conclude that a four-body encounter cannot
satisfactorily explain the observed velocities of VFTS~590 and
Mel~34, nor the orbital separation of the latter.

\subsection{5-body scattering}\label{Sect:5body_scattering}

The most elaborate and complicated calculations are the 5-body
encounters, for which we also use the {\sc sigma} package
\cite{2004MNRAS.350..615G} from {\sc starlab}
\cite{2004MNRAS.351..473P}. For those, we focus on a target triple
((\tAa,\tAb),\tB)) that interacts with a projectile binary (\pA, \pB).

We performed a total of 1,565,520 5-body scattering experiments
distributed over 76 cross-section calculations for a range of initial
conditions.  With the calculations, we pin-point the choice of initial
conditions calibrated for the theoretical arguments set out in main
manuscript. These arguments are based on the conservation of energy
before and after the dynamical interaction, and the expected amount of
energy released in the interaction (see \cref{Eq:dEbin}).  We further
explore, as a free parameters, the relative encounter velocity
($v_{\rm enc}$) between 1\,km/s and 13\,km/s (see
\cref{tab:cross_sections_velocity_dependence}), and the orbit of the
tertiary star (which is ill constrained by the energetics, see
\cref{tab:cross_sections_orbital_separation}).  Once the optimal set
of parameters for the pre-encounter configuration of the binary and
the triple are found, we perform performed an extra set of runs to
study the sensitivity of the conditions.  We perfomed 76 cross section
calculations for 5-body (binary-triple) interactions.

After constraining the parameters for the 5-body encounters, we focus
on the interaction between a target triple and a projectile binary. To
limit parameter space, we assume that VFTS~590 has a mass of
50\,\MSun\, Mel39B is 80\,\MSun\, and the other stars are 130\,\MSun.
We realize that the slight differences in mass between Mel~34A
(139\,\MSun), Mel~34B (127\,\MSun), and Mel~39A (140\,\MSun) are
relevant for our calculations, but the simplification helps in
acquiring a global understanding of the consequences of the
interaction, before constraining the specific masses of Mel34A,
Mel34B, and Mel39A, which are rather similar in a dynamical
perspective.

After determining the cross sections, we realize scenario G is
most promising, as explained in the main text. In order to explore the
effect of the incoming relative velocity, we adopt model G and
calculate the interaction with a relative velocities of 1\,km/s to
13\,km/s. The results of these calculations are presented in
\cref{tab:cross_sections_velocity_dependence}.

To illustrate the sensitivity of the orbital separation for the two
binaries, we present the results of cross-section calculations for
slightly wider systems. We rather arbitrarily adopted 10\,au for the
inner target binary, 100\,au for the orbital separation of the
tertiary component, and 10\,au for the projectile binary. The
resulting cross-sections for the 7 configurations in which we can
distribute VTS~590 and Mel39B over the triple and binary are
presented in\,\cref{tab:cross_sections_for_1010010}.

\begin{figure}
\center
\includegraphics[width=1.0\textwidth]{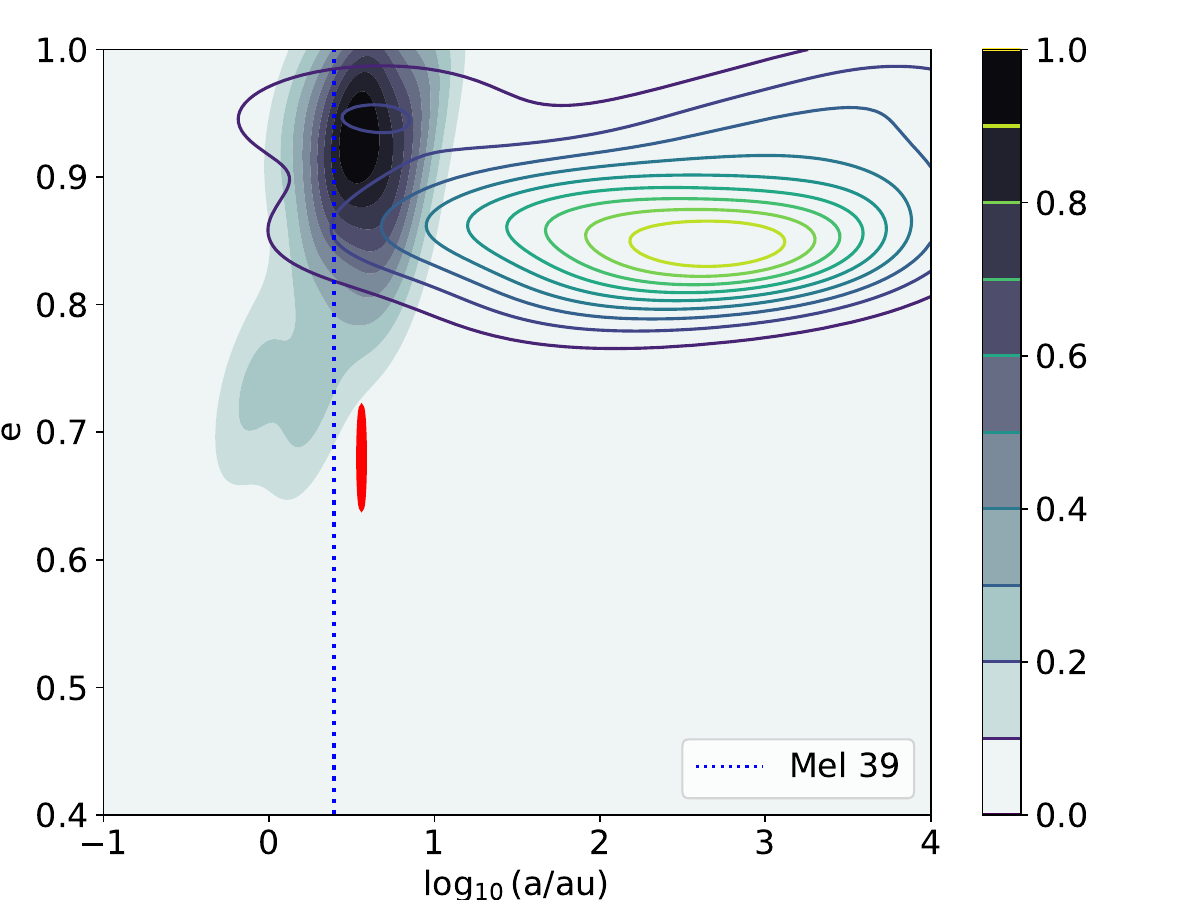}
\caption[]{Orbital separation vs eccentricity for scenario G in which
  Mel~34 (grey shades, color bar to the right) and Mel~39 (contours,
  overplotted in the right-hand color bar) for an interaction in which
  the inner target and the projectile binaries have an orbital
  separation of 10\,au, whereas the tertiary star (VFTS~590) is
  100\,au. Note the four-times wider range in the x-axis compared to
  \cref{fig:sma_vs_ecc_Mel34_Mel39}.  The red ellipsoid point gives
  the observed orbital parameters for Mel~34 (its size is 4 standard
  deviations, twice the range as in
  \cref{fig:sma_vs_ecc_Mel34_Mel39}), and the vertical blue dotted
  line gives the observed semi-major axis (assuming an 80\,\MSun\,
  star for the secondary) for Mel~39.}
\label{fig:sma_vs_ecc_1010010}
\end{figure}

In figure\,\ref{fig:sma_vs_ecc_1010010} we present, for model G of
\cref{tab:cross_sections_for_1010010}, the orbital separation and
eccentricity for the runaway binaries for Mel~34, Mel~39 after an
interaction with initial (pre-encounter) orbital parameters somewhat
wider than adopted in the main paper. Comparing
figure\,\ref{fig:sma_vs_ecc_1010010} with
figure\,\ref{fig:sma_vs_ecc_Mel34_Mel39} illustrates the sensitivity
of the initial orbital parameters to the relative orbital
distributions (in semi-major axis and eccentricity) of the final
runaway binaries.

The relative probabilities for the 8 configurations for a
triple-binary encounter are presented in \cref{tab:cross_sections}.
We further specify the interaction by calculating the relative
frequency of each of the various branching ratios for the interactions
for scenarios A to H (see table\,\ref{tab:interaction_frequency}
in~\cref{sect:crosssections}). The encounters for model G (where
VFTS~590 orbiting the target binary) is most probable, leading to a
rate of $\sim 27$\,Myr$^{-1}$, which is conveniently close to the
observed rate of $24$\,Myr$^{-1}$ \citep[see][]{2024Natur.634..809S}. The most
probable interaction (in model G) is the preservation of both binaries
(see\,\cref{tab:cross_sections}). .

It may seem that we rather quickly focus on the encounter between a
4\,au\, and 6\,au binary, but many parameters were varied to constrain
these parameters. We explored the inner target and projectile orbital
separation between 2\,au and 10\,au (with constant increments of
1\,au), and the tertiary orbit between 30\,au and 320\,au (with
constant increments of 10\,au).  The upper limit corresponds to about
half the semi-major axis for a hard-soft encounter in the R136's
cluster core. The relative velocities ware varied between 1\,km/s and
13\,km/s (with constant increments of 1\,km/s).  Note that stars with
a velocities $>7$\,km/s have escape speed from the cluster core
region.

\begin{table}
\begin{tabular}{lcrrrrr}
\hline
id & escaper    & (\pA, \pB) & (\pA, \tAa) & (\pA, \tB) & (\pA, \tAb) \\
   & (VFTS~590) \\
\hline
A & \tB & 0 & 0 & -- & 0 \\
B & \tB & 0.61 & 0.15 & -- & 0.11 \\
C & \pB & -- & 0 & 0 & 0 \\
D & \tAb& 0.75 & 0.25 & 0 & -- \\
E & \tB & 0 & 0 & -- & 0 \\
F & \tAb& 0.54 & 0.29 &0.18& -- \\
G & \tB & 0.83 & 0.17 & -- & 0.06 \\
H & \tAb& 0.75 & 0.17 &0.08& -- \\

\hline
\end{tabular}
\caption[]{Identification of the escaping star and relative outcome
frequency for scenarios A till H for the encounter in which $a_{\rm
in} = 4$\,au, $a_{\rm out} = 40$\,au, $a_{\rm projectile} =
6$\,au. For an encounter velocity of $v_{\rm enc} = 7$\,km/s.
Scenario G in which \tB\, escapes as a single and (\pA, \pB) as well
as (\tAa, \tAb) escape as binaries give the most probable outcome.}
\label{tab:interaction_frequency}
\end{table}

\begin{table}
\begin{tabular}{lllllllll}
\hline
$v_{\rm enc}$ & $\sigma$ & $f_{\rm esc, VFTS590}$& $\sigma_{\rm run}$ &$ \Gamma_{\rm esc, VFTS590}$ & (\pA, \pB) & (\pA, \tAa) & (\pA, \tAb) \\
$$[km/s] & [$10^6$ au$^2$] & & [$10^6$ au$^2$] & [Myr$^{-1}$] \\
\hline
1& 34700 & 0.348 & 0.053 &1840 & 0.85 & 0.95 & 0.05 \\
3& 1280 & 0.383 & 0.040 & 156 & 0.81 & 0.13 & 0.06 \\
5& 277 & 0.358 & 0.028 & 38.5 & 0.73 & 0.18 & 0.09 \\
7& 101 & 0.323 & 0.048 & 33.9 & 0.83 & 0.17 & 0.06 \\
9& 47.6 & 0.340 & 0.045 & 19.5 & 0.72 & 0.17 & 0.11 \\
11& 26.1 & 0.340 & 0.038 & 10.9 & 0.73 & 0.13 & 0.13 \\
13& 15.8 & 0.365 & 0.045 & 9.33& 0.94 & 0.06 & 0.00 \\
\hline
\end{tabular}
\caption[]{Cross section for scenario G but with varying the encounter
velocity. Here we adopted the binary (\tAa,\tAb) to have a
semi-major axis of $a_{\rm in} =4$\,au, The initial tertiary
\tB\, (that eventually escapes) orbits the target binary with
a semimajor-axis of $a_{\rm out} = 40$\,au. The projectile binary
(\pA, \pB), has a semi-major axis of 6\,au. The first column gives
the relative encounter velocity, followed by the cross section, and
the fraction of systems for which VFTS~590 escapes. The following
columns give the cross section for VFTS~590 and the two binaries to
acquire a minimum runaway velocity of 27.6km/s, and the encounter
rate associated with this cross section. The last three columns give
the branching ratios of the eventual surviving binaries with \pA\,
as primary. }
\label{tab:cross_sections_velocity_dependence}
\end{table}

Eventually, we adopt the most opportune result from the relative
encounter velocity, and vary the orbital separation of the tertiary
star (VFTS~590). The results are presented in
table\,\ref{tab:cross_sections_orbital_separation}. The orbital
separation of the tertiary star has only a minor effect on the
interaction cross-section, and on the other final orbital parameters
of the two runaway binaries and the velocity of all three runaways. It
therefore remains hard to constrain the pre-encounter orbit of
VFTS~590 around the binary Mel~39.

\begin{table}
\begin{tabular}{rrcccrrr}
\hline
$a_{\rm out}$ & $\sigma$ & $f_{\rm esc, VFTS590}$& $\sigma_{\rm run}$ & $\Gamma_{\rm esc, VFTS590}$ & (\pA, \pB) & (\pA, \tAa) & (\pA, \tAb) \\
& & & & & (\tAa, \tAb) & (\pB, \tAb) & (\pB, \tAa) \\
$$[au] & [$10^6$ au$^2$] & [au$^2$] & [au$^2$] & [Myr]$^{-1}$ \\
\hline
40 & 103 & 0.43 & 0.046 & 34 & 0.83 & 0.17 & 0.06\\
50 & 142 & 0.42 & 0.043 & 43 & 0.77 & 0.09 & 0.14\\
60 & 140 & 0.43 & 0.040 & 39 & 0.80 & 0.11 & 0.10\\
80 & 99 & 0.28 & 0.033 & 23 & 0.75 & 0.16 & 0.09\\
100& 135 & 0.45 & 0.043 & 41 & 0.73 & 0.14 & 0.13\\
120& 95 & 0.56 & 0.049 & 33 & 0.72 & 0.14 & 0.13\\
140& 134 & 0.51 & 0.042 & 39 & 0.77 & 0.12 & 0.11\\
160& 133 & 0.28 & 0.026 & 24 & 0.81 & 0.06 & 0.13\\
180& 132 & 0.65 & 0.038 & 36 & 0.83 & 0.08 & 0.09\\
200& 132 & 0.68 & 0.039 & 36 & 0.75 & 0.10 & 0.15\\
240& 132 & 0.80 & 0.023 & 21 & 0.75 & 0.13 & 0.12\\
\hline
\end{tabular}
\caption[]{Branching ratios and cross-section for scenario G (\tB\,
escapes) for the varying orbit of the tertiary star (VFTS~590) around
the inner binary (\tAa, \tAb). The inner target binary has an
adopted semi-major axis of 4\,au, while the projectile binary has a
semi-major axis of 6\,au. The relative encounter velocity is
7\,km/s. The first column gives the semi-major axis of the tertiary
star (\tB) around the target binary (\tAa, \tAb). The following
column gives the cross section for this interaction to happen. The
next three columns give the fraction of systems, cross-section and
the rate for the escaper (VFTS~590) to acquire a runaway speed of at
least 27.6\,km/s. The last three columns give the branching ratios
of the eventual surviving binaries with \pA\, as primary. }
\label{tab:cross_sections_orbital_separation}
\end{table}

\begin{table}
\begin{tabular}{clccrc}
\hline id & Encounter & $\sigma$ & $f_{\rm esc, VFTS590}$& $f_{\rm run}$ &$ \Gamma_{\rm esc, VFTS590}$ \\
& & [$10^6$ au$^2$] & [au$^2$] & & [Myr]$^{-1}$ \\
\hline
A & $\langle$((\tAa, \tAb), \tB), (Mel39B, VFTS~590)$\rangle$ & 132 & 0.003 & 0.000 & 0.00 \\
B & $\langle$((\tAa, \tAb), VFTS~590), (\pA, Mel39B)$\rangle$ & 144 & 0.502 & 0.008 & 7.56 \\
C & $\langle$((\tAa, \tAb), Mel39B), (\pA, VFTS~590)$\rangle$ & 134 & 0.001 & 0.000 & 1.25 \\
D & $\langle$((\tAa, VFTS~590), \tB), (\pA, Mel39B)$\rangle$ & 231 & 0.174 & 0.002 & 284 \\
E & $\langle$((\tAa, Mel39B), \tB), (\pA, VFTS~590)$\rangle$ & 148 & 0.002 & 0.000 & 1.62 \\
F & $\langle$((\tAa, VFTS~590), Mel39B), (\pA, \pB)$\rangle$ & 290 & 0.128 & 0.003 & 263 \\
G & $\langle$((\tAa, Mel39B), VFTS~590), (\pA, \pB)$\rangle$ & 281 & 0.492 & 0.042 & 977 \\
H & $\langle$((Mel39B, VFTS~590), \tB), (\pA, \pB)$\rangle$ & 296 & 0.141 & 0.004 & 296 \\
\hline
\end{tabular}
\caption[]{Cross sections for the 8 reduced cases for which we adopt
Mel39B as a 80\,\MSun\, star, and 50\,\MSun\, for VFTS~590. All the
other stars are 140\,\MSun. Calculations are performed with a
semi-major axis for the inner target binary of $a = 10$\,au and an
outer orbit of $100$\,au. For the projectile binary, we also adopt
$10$\,au. The relative encounter velocity is $v_{\rm enc} =
7$\,km/s. The first column identifies the scenario, explains the
second column. The third and fourth columns give the total cross
section for this interaction and the fraction of cases in which
VFTS~590 is ejected. The last two columns give the cross-section
and rate for VFTS~590 to be ejected with a velocity of at least
27.6km/s. }
\label{tab:cross_sections_for_1010010}
\end{table}

\section{The evolution of the runaway binaries}\label{SM:Binev}

After being ejected from the R136 cluster, the two binaries, Mel~34
and Mel~39, continue to evolve, and Roche-lobe overflow is inevitable
for both. We continue their evolution assuming that they were not
further perturbed by encounters with other stars.  We use the binary
evolution code SeBa \cite{1996A&A...309..179P} to estimate the
evolution of the two binaries, assuming that they were ejected at an
age of 1\,Myr, and that they remain unperturbed for the rest of their
lifetimes.

\begin{figure}
\center
\includegraphics[width=1.0\textwidth]{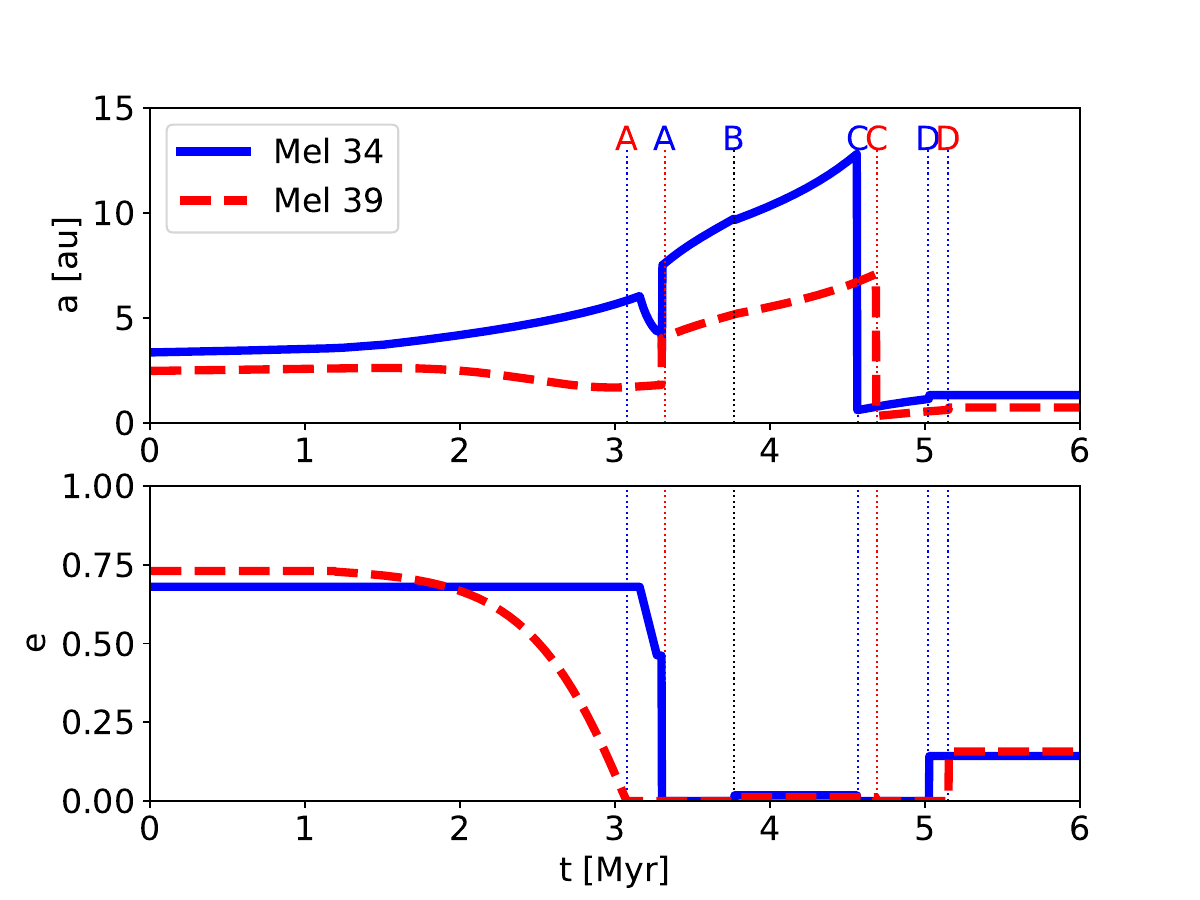}
\caption[]{Orbital evolution of the two binaries Mel~34 (blue) and
  Mel~39 (red) after they have been ejected from the star cluster
  R136. The top panel shows the evolution of the orbital separation,
  the bottom planet shows the eccentricity. We identify 4 distinct
  events in the binary evolution, which are A: the onset of mass
  transfer, B: the first supernova, C: the initiation of the second
  phase of mass transfer, and D: the second supernova at about 6\,Myr
  (after which we stop the calculation). The letters correspond to the
  short-hand notation presented in the text.  Both binaries remain
  bound afterwards, but eventually (after many Hubble times) will
  merge due to the emission of gravitational waves.
\label{fig:Mel34_and_Mel39_evolution}
}
\end{figure}

Both binaries, Mel~34 and Mel~39 have quite a comparable evolution,
which starts with two main-sequence stars in rather tight and elliptic
orbits.  We illustrate the evolution of both binaries in
\cref{fig:Mel34_and_Mel39_evolution}, and indicate a few key phases by
the letters A to D.  The shorthand notation notation, introduced in
\cite{1996A&A...309..179P} can be written as

  \begin{tabular}{ccccccccccccccccccc}
    && &&A&& &&B&&C&& &&D&& && \\
(ms, ms) &$\rightarrow$& (gs, ms)$_{\rm c}$ &$\rightarrow$& [gs, ms) &$\rightarrow$& (WR, ms) &$\rightarrow$& (bh, ms) &$\rightarrow$&
  (bh, gs]$_{\rm c}$ &$\rightarrow$& \{bh, gs\}     &$\rightarrow$& (bh, WR) &$\rightarrow$& (bh, bh) &$\rightarrow$& \{bh\}. \\
  \end{tabular}
  \bigskip

The evolution starts with two main sequence stars, indicates as $(ms,
ms)$ in elliptic orbits (see lower panel in
\cref{fig:Mel34_and_Mel39_evolution}).  In the coming 3\,Myr the
orbits of both binaries circularizes (indicated with the subscript c),
and the stellar winds causes the orbits to widen.  The primary star
(indicated in the left hand side) ascends the giant branch (gs) just
before its hydrogen envelope runs out, and fills it's Roche lobe
(indicated in \cref{fig:Mel34_and_Mel39_evolution} with the vertical
dashed line and the letter A).  Mass transfer from the primary star
(The Roche-lobe filling star is indicated in the short-hand notation
with the square bracket) to the secondary. The primary, being stripped
of it's envelope, becomes a naked helium star (WR) and collapses a
some time later to a black hole (bh, indicated with the letter B in
\cref{fig:Mel34_and_Mel39_evolution}).  The first supernovae in Mel~34
and Mel~39 occur around the same time, and for clarity, is only
indicated in blue in \cref{fig:Mel34_and_Mel39_evolution}.  The
secondary star subsequently, ascends the giant branch, circularizes
the orbit, and commences into a common envelope (indicated with
braces, and the letter C in \cref{fig:Mel34_and_Mel39_evolution}). The
resulting tight binary is composed of the primary black hole and the
stripped core of the Roche-lobe filling giant. The resulting
Wolf-Rayet star collapses into a black hole in a supernova (indicated
by the letter D in \cref{fig:Mel34_and_Mel39_evolution}). If the
binary survives (here we neglected natal kicks in the supernova,
allowing both binaries to survive the mass loss in the supernovae) the
two black holes eventually merge into a single black hole due to the
emission of gravitational waves.

The orbital separation of the two black hole binaries, however, is too
large, and their eccentricities too small to make the black holes
merge in a Hubble time due to the emission of gravitational waves.

\section*{Energy consumption of this calculation}

In the spirit of the energy consumption of our scientific endeavour
\cite{2020NatAs...4..819P}, we report on our energy consumption for
the calculations in this research. The calculations using {\sc Starlab} and
AMUSE took about $2.9\cdot10^7$ core seconds. Runs are performed on a
165\,Watt 12-core Xeon E-2176M CPU with NVIDIA Quadro P1000 Max-Q GPU,
totaling about 110\,KWh. The origin of the electricity was wind and
solar.



\begin{thebibliography}{10}
\expandafter\ifx\csname url\endcsname\relax
  \def\url#1{\texttt{#1}}\fi
\expandafter\ifx\csname urlprefix\endcsname\relax\def\urlprefix{URL }\fi
\expandafter\ifx\csname href\endcsname\relax
  \def\href#1#2{#2} \def\path#1{#1}\fi

\bibitem{2024Natur.634..809S}
M.~{Stoop}, A.~{de Koter}, L.~{Kaper}, S.~{Brands}, S.~{Portegies Zwart},
  H.~{Sana}, F.~{Stoppa}, M.~{Gieles}, L.~{Mahy}, T.~{Shenar}, D.~{Guo},
  G.~{Nelemans}, S.~{Rieder}, {Two waves of massive stars running away from the
  young cluster R136}, \nat 634~(8035) (2024) 809--812.
\newblock \href {http://arxiv.org/abs/2410.06255} {\path{arXiv:2410.06255}},
  \href {http://dx.doi.org/10.1038/s41586-024-08013-8}
  {\path{doi:10.1038/s41586-024-08013-8}}.

\bibitem{1961BAN....15..265B}
A.~{Blaauw}, {On the origin of the O- and B-type stars with high velocities
  (the "run-away" stars), and some related problems}, \bain 15 (1961) 265--+.

\bibitem{1961BAN....15..291B}
J.~{Boersma}, {Mathematical theory of the two-body problem with one of the
  masses decreasing with time}, \bain 15 (1961) 291--301.

\bibitem{1983ApJ...267..322H}
J.~G. {Hills}, {The effects of sudden mass loss and a random kick velocity
  produced in a supernova explosion on the dynamics of a binary star of
  arbitrary orbital eccentricity - Applications to X-ray binaries and to the
  binary pulsars}, \apj 267 (1983) 322--333.
\newblock \href {http://dx.doi.org/10.1086/160871} {\path{doi:10.1086/160871}}.

\bibitem{2011Sci...334.1380F}
M.~S. {Fujii}, S.~{Portegies Zwart}, {The Origin of OB Runaway Stars}, Science
  334 (2011) 1380--.
\newblock \href {http://arxiv.org/abs/1111.3644} {\path{arXiv:1111.3644}},
  \href {http://dx.doi.org/10.1126/science.1211927}
  {\path{doi:10.1126/science.1211927}}.

\bibitem{2010ARA&A..48..431P}
S.~F. {Portegies Zwart}, S.~L.~W. {McMillan}, M.~{Gieles}, {Young Massive Star
  Clusters}, \araa 48 (2010) 431--493.
\newblock \href {http://arxiv.org/abs/1002.1961} {\path{arXiv:1002.1961}},
  \href {http://dx.doi.org/10.1146/annurev-astro-081309-130834}
  {\path{doi:10.1146/annurev-astro-081309-130834}}.

\bibitem{2010MNRAS.408..731C}
P.~A. {Crowther}, O.~{Schnurr}, R.~{Hirschi}, N.~{Yusof}, R.~J. {Parker}, S.~P.
  {Goodwin}, H.~A. {Kassim}, {The R136 star cluster hosts several stars whose
  individual masses greatly exceed the accepted 150M$_{solar}$ stellar mass
  limit}, \mnras 408~(2) (2010) 731--751.
\newblock \href {http://arxiv.org/abs/1007.3284} {\path{arXiv:1007.3284}},
  \href {http://dx.doi.org/10.1111/j.1365-2966.2010.17167.x}
  {\path{doi:10.1111/j.1365-2966.2010.17167.x}}.

\bibitem{2022A&A...663A..36B}
S.~A. {Brands}, A.~{de Koter}, J.~M. {Bestenlehner}, P.~A. {Crowther}, J.~O.
  {Sundqvist}, J.~{Puls}, S.~M. {Caballero-Nieves}, M.~{Abdul-Masih}, F.~A.
  {Driessen}, M.~{Garc{\'\i}a}, S.~{Geen}, G.~{Gr{\"a}fener}, C.~{Hawcroft},
  L.~{Kaper}, Z.~{Keszthelyi}, N.~{Langer}, H.~{Sana}, F.~R.~N. {Schneider},
  T.~{Shenar}, J.~S. {Vink}, {The R136 star cluster dissected with Hubble Space
  Telescope/STIS. III. The most massive stars and their clumped winds}, \aap
  663 (2022) A36.
\newblock \href {http://arxiv.org/abs/2202.11080} {\path{arXiv:2202.11080}},
  \href {http://dx.doi.org/10.1051/0004-6361/202142742}
  {\path{doi:10.1051/0004-6361/202142742}}.

\bibitem{2007Natur.450..388P}
S.~F. {Portegies Zwart}, E.~P.~J. {van den Heuvel}, {A runaway collision in a
  young star cluster as the origin of the brightest supernova}, \nat 450 (2007)
  388--389.
\newblock \href {http://arxiv.org/abs/arXiv:0711.2293}
  {\path{arXiv:arXiv:0711.2293}}, \href {http://dx.doi.org/10.1038/nature06276}
  {\path{doi:10.1038/nature06276}}.

\bibitem{1954ApJ...119..625B}
A.~{Blaauw}, W.~W. {Morgan}, The space motions of ae aurigae and \&mu; columbae
  with respect to the orion nebula., \apj 119 (1954) 625.

\bibitem{Poveda_1967}
A.~{Poveda}, J.~{Ruiz}, C.~{Allen}, Bolet\'{\i}n de los Observatorios de
  Tonantzintla y Tacubaya.

\bibitem{2024A&A...690A.207P}
B.~{Polak}, M.-M. {Mac Low}, R.~S. {Klessen}, S.~{Portegies Zwart}, E.~P.
  {Andersson}, S.~M. {Appel}, C.~{Cournoyer-Cloutier}, S.~C.~O. {Glover},
  S.~L.~W. {McMillan}, {Massive star cluster formation: II. Runaway stars as
  fossils of subcluster mergers}, \aap 690 (2024) A207.
\newblock \href {http://arxiv.org/abs/2405.12286} {\path{arXiv:2405.12286}},
  \href {http://dx.doi.org/10.1051/0004-6361/202450774}
  {\path{doi:10.1051/0004-6361/202450774}}.

\bibitem{2019MNRAS.484.2692T}
K.~A. {Tehrani}, P.~A. {Crowther}, J.~M. {Bestenlehner}, S.~P. {Littlefair},
  A.~M.~T. {Pollock}, R.~J. {Parker}, O.~{Schnurr}, {Weighing Melnick 34: the
  most massive binary system known}, \mnras 484~(2) (2019) 2692--2710.
\newblock \href {http://arxiv.org/abs/1901.04769} {\path{arXiv:1901.04769}},
  \href {http://dx.doi.org/10.1093/mnras/stz147}
  {\path{doi:10.1093/mnras/stz147}}.

\bibitem{2000IJoMP...15..4871P}
S.~F. {Portegies Zwart}, S.~L.~W. {McMillan},
  \href{http://journals.wspc.com.sg/139/15/1530/S0217751X00002135.html}{{Gravitational
  thermodynamics and black-hole mergers}}, Int, J, of Mod. Phys, A 15 (2000)
  4871--4875.
\newblock \href {http://dx.doi.org/10.1142/S0217751X00002135}
  {\path{doi:10.1142/S0217751X00002135}}.
\newline\urlprefix\url{http://journals.wspc.com.sg/139/15/1530/S0217751X00002135.html}

\bibitem{1975MNRAS.173..729H}
D.~C. {Heggie}, Binary evolution in stellar dynamics, \mnras 173 (1975)
  729--787.

\bibitem{2003gmbp.book.....H}
D.~{Heggie}, P.~{Hut}, {The Gravitational Million-Body Problem: A
  Multidisciplinary Approach to Star Cluster Dynamics}, The Gravitational
  Million-Body Problem: A Multidisciplinary Approach to Star Cluster Dynamics,
  by Douglas Heggie and Piet Hut.~ Cambridge University Press, 2003, 372 pp.,
  2003.

\bibitem{1992ApJ...389..527H}
P.~{Hut}, S.~{McMillan}, R.~W. {Romani}, {The evolution of a primordial binary
  population in a globular cluster}, \apj 389 (1992) 527--545.

\bibitem{2002ApJ...576..899P}
S.~F. {Portegies Zwart}, S.~L.~W. {McMillan}, {The Runaway Growth of
  Intermediate-Mass Black Holes in Dense Star Clusters}, \apj 576 (2002)
  899--907.
\newblock \href {http://arxiv.org/abs/astro-ph/0201055}
  {\path{arXiv:astro-ph/0201055}}, \href {http://dx.doi.org/10.1086/341798}
  {\path{doi:10.1086/341798}}.

\bibitem{2020MNRAS.499.1918B}
J.~M. {Bestenlehner}, P.~A. {Crowther}, S.~M. {Caballero-Nieves}, F.~R.~N.
  {Schneider}, S.~{Sim{\'o}n-D{\'\i}az}, S.~A. {Brands}, A.~{de Koter},
  G.~{Gr{\"a}fener}, A.~{Herrero}, N.~{Langer}, D.~J. {Lennon}, J.~{Ma{\'\i}z
  Apell{\'a}niz}, J.~{Puls}, J.~S. {Vink}, {The R136 star cluster dissected
  with Hubble Space Telescope/STIS - II. Physical properties of the most
  massive stars in R136}, \mnras 499~(2) (2020) 1918--1936.
\newblock \href {http://arxiv.org/abs/2009.05136} {\path{arXiv:2009.05136}},
  \href {http://dx.doi.org/10.1093/mnras/staa2801}
  {\path{doi:10.1093/mnras/staa2801}}.

\bibitem{2025MNRAS.539.1291P}
A.~M.~T. {Pollock}, P.~A. {Crowther}, J.~M. {Bestenlehner}, P.~S. {Broos},
  L.~K. {Townsley}, {Melnick 39 is a very massive intermediate-period
  colliding-wind binary}, \mnras 539~(2) (2025) 1291--1298.
\newblock \href {http://arxiv.org/abs/2503.17150} {\path{arXiv:2503.17150}},
  \href {http://dx.doi.org/10.1093/mnras/staf501}
  {\path{doi:10.1093/mnras/staf501}}.

\bibitem{2001MNRAS.321..398M}
R.~A. {Mardling}, S.~J. {Aarseth}, {Tidal interactions in star cluster
  simulations}, \mnras 321 (2001) 398--420.

\bibitem{1977A&A....57..383Z}
J.-P. {Zahn}, {Tidal friction in close binary stars}, \aap 57 (1977) 383--394.

\bibitem{2002ApJ...565..982M}
P.~{Massey}, L.~R. {Penny}, J.~{Vukovich}, {Orbits of Four Very Massive
  Binaries in the R136 Cluster}, \apj 565 (2002) 982--993.
\newblock \href {http://arxiv.org/abs/arXiv:astro-ph/0110088}
  {\path{arXiv:arXiv:astro-ph/0110088}}, \href
  {http://dx.doi.org/10.1086/324783} {\path{doi:10.1086/324783}}.

\bibitem{2018LRR....21....3A}
B.~P. {Abbott},  et al.,
  {Kagra Collaboration}, {VIRGO Collaboration}, {Prospects for observing and
  localizing gravitational-wave transients with Advanced LIGO, Advanced Virgo
  and KAGRA}, Living Reviews in Relativity 21~(1) (2018) 3.
\newblock \href {http://arxiv.org/abs/1304.0670} {\path{arXiv:1304.0670}},
  \href {http://dx.doi.org/10.1007/s41114-018-0012-9}
  {\path{doi:10.1007/s41114-018-0012-9}}.

\bibitem{10.1088/978-0-7503-1320-9}
S.~Portegies~Zwart, S.~McMillan,
  \href{http://dx.doi.org/10.1088/978-0-7503-1320-9}{Astrophysical Recipes},
  2514-3433, IOP Publishing, 2018.
\newblock \href {http://dx.doi.org/10.1088/978-0-7503-1320-9}
  {\path{doi:10.1088/978-0-7503-1320-9}}.
\newline\urlprefix\url{http://dx.doi.org/10.1088/978-0-7503-1320-9}

\bibitem{1955ApJ...121..161S}
E.~E. {Salpeter}, The luminosity function and stellar evolution., \apj 121
  (1955) 161.

\bibitem{1911MNRAS..71..460P}
H.~C. {Plummer}, {On the problem of distribution in globular star clusters},
  \mnras 71 (1911) 460--470.

\bibitem{2014MNRAS.445..674C}
D.~P. {Caputo}, N.~{de Vries}, S.~{Portegies Zwart}, {On the effects of
  subvirial initial conditions and the birth temperature of R136}, \mnras
  445~(1) (2014) 674--685.
\newblock \href {http://arxiv.org/abs/1409.4765} {\path{arXiv:1409.4765}},
  \href {http://dx.doi.org/10.1093/mnras/stu1769}
  {\path{doi:10.1093/mnras/stu1769}}.

\bibitem{1966AJ.....71...64K}
I.~R. {King}, The structure of star clusters. iii. some simple dvriamical
  models, \aj 71 (1966) 64--75.

\bibitem{1991ApJ...383..181M}
J.~{Makino}, P.~{Hut}, {On core collapse}, \apj 383 (1991) 181--191.
\newblock \href {http://dx.doi.org/10.1086/170774} {\path{doi:10.1086/170774}}.

\bibitem{2009MNRAS.396..570G}
V.~V. {Gvaramadze}, A.~{Gualandris}, S.~{Portegies Zwart}, {On the origin of
  high-velocity runaway stars}, \mnras 396 (2009) 570--578.
\newblock \href {http://arxiv.org/abs/0903.0738} {\path{arXiv:0903.0738}},
  \href {http://dx.doi.org/10.1111/j.1365-2966.2009.14809.x}
  {\path{doi:10.1111/j.1365-2966.2009.14809.x}}.

\bibitem{2004MNRAS.351..473P}
S.~F. {Portegies Zwart}, P.~{Hut}, S.~L.~W. {McMillan}, J.~{Makino}, {Star
  cluster ecology - V. Dissection of an open star cluster: spectroscopy},
  \mnras 351 (2004) 473--486.
\newblock \href {http://arxiv.org/abs/arXiv:astro-ph/0301041}
  {\path{arXiv:arXiv:astro-ph/0301041}}, \href
  {http://dx.doi.org/10.1111/j.1365-2966.2004.07709.x}
  {\path{doi:10.1111/j.1365-2966.2004.07709.x}}.

\bibitem{2004MNRAS.350..615G}
A.~{Gualandris}, S.~{Portegies Zwart}, P.~P. {Eggleton}, {N-body simulations of
  stars escaping from the Orion nebula}, \mnras 350 (2004) 615--626.
\newblock \href {http://arxiv.org/abs/arXiv:astro-ph/0401451}
  {\path{arXiv:arXiv:astro-ph/0401451}}, \href
  {http://dx.doi.org/10.1111/j.1365-2966.2004.07673.x}
  {\path{doi:10.1111/j.1365-2966.2004.07673.x}}.

\bibitem{1986LNP...267..233H}
D.~C. {Heggie}, R.~D. {Mathieu}, {Standardised Units and Time Scales}, in:
  P.~{Hut}, S.~L.~W. {McMillan} (Eds.), The Use of Supercomputers in Stellar
  Dynamics, Vol. 267 of Lecture Notes in Physics, Berlin Springer Verlag, 1986,
  p. 233.
\newblock \href {http://dx.doi.org/10.1007/BFb0116419}
  {\path{doi:10.1007/BFb0116419}}.

\bibitem{1996A&A...309..179P}
S.~F. {Portegies Zwart}, F.~{Verbunt}, {Population synthesis of high-mass
  binaries.}, \aap 309 (1996) 179--196.

\bibitem{2020NatAs...4..819P}
S.~{Portegies Zwart}, {The ecological impact of high-performance computing in
  astrophysics}, Nature Astronomy 4 (2020) 819--822.
\newblock \href {http://arxiv.org/abs/2009.11295} {\path{arXiv:2009.11295}},
  \href {http://dx.doi.org/10.1038/s41550-020-1208-y}
  {\path{doi:10.1038/s41550-020-1208-y}}.

\end{thebibliography}
\end{appendix}

\end{document}